# Distant activity of comet C/2002 VQ94 (LINEAR): optical spectrophotometric monitoring between 8.4 and 16.8 au from the Sun

Pavlo P. Korsun[a,*], Philippe Rousselot[b], Irina V. Kulyk[a], Viktor L. Afanasiev[c], Oleksandra V. Ivanova[a]

[a] *Main Astronomical Observatory of NAS of Ukraine, Akademika Zabolotnoho 27, 03680 Kyiv, Ukraine*
[*] *Corresponding Author. E-mail address:* korsun@mao.kiev.ua

[b] *University of Franche-Comté, Observatoire des Sciences de l'Univers THETA, Institut UTINAM - UMR CNRS 6213, BP 1615, 25010 Besançon Cedex, France*

[c] *Special Astrophysical Observatory of the Russian AS, Nizhnij Arkhyz, 369167, Russia*

**Abstract**

Spectrophotometric monitoring of distant comet C/2002 VQ94 (LINEAR) was performed with the 6-m telescope of SAO RAS (Special Astrophysical Observatory of Russian Academy of Sciences) and with the 2.5-m Nordic Optical Telescope (Observatory del Roque de los Muchachos, Canarias, Spain) between 2008 and 2013. During this period the comet was on the outbound segment of its orbit, between heliocentric distances of 8.36 au and 16.84 au. Analysis of the spectra revealed the presence of the $CO^+$ and $N_2^+$ emissions in the cometary coma at a distance of 8.36 au from the Sun. This distance is larger than ionic emissions have been detected in any previous objects. Only continuum, with no traces of emissions, was detected in the spectrum obtained in 2009 when the comet was at a distance of 9.86 au. From the spectra obtained in 2008, average column densities of $2.04 \times 10^9$ mol cm$^{-2}$ for $N_2^+$ and $3.26 \times 10^{10}$ mol cm$^{-2}$ for $CO^+$ were measured in the cometary coma. The derived values correspond to $N_2^+/CO^+=0.06$ within the projected slit. Images obtained through a red continuum filter in 2008 showed a bright, dust coma, indicating a high level of physical activity. A considerably lower level of activity was observed in 2009 and 2011 at distances of 9.86 au and 13.40 au respectively. No noticeable activity was detected in 2013 at a heliocentric distance of 16.84 au. The $Af\rho$ parameter, which is used as an indicator of cometary activity, was measured as 2000 cm in 2008, and 800 cm in 2009 and 2011. The $Af\rho$ values correspond to dust production rates between 10-20 kg s$^{-1}$, 4-6 kg s$^{-1}$ and 3-5 kg s$^{-1}$ at 8.36, 9.86, and 13.40 au respectively. There is an obvious correlation between the decrease of the dust production rate of the nucleus and the disappearance of the emissions in the spectrum of C/2002 VQ94 (LINEAR) at heliocentric distances greater than 9 au. The colors and size of the nucleus of C/2002 VQ94 (LINEAR) were estimated from the images obtained during the late stage at a heliocentric distance of 16.84 au, when the activity had probable ceased. The B-V and V-R colors were estimated to be 1.07±0.05 and 0.54±0.03 respectively. The effective nucleus radius of 48±2 km is in agreement with the previously published results, obtained from the observations of the comet during its early inactive stage (Jewitt, 2005; Weiler et al., 2011).

## 1. Introduction

Comet C/2002 VQ94 (LINEAR) (hereafter VQ94) was discovered by the LINEAR team on 2002 November 11.24 UT at a distance of 10.02 au from the Sun (Tichy et al., 2002). It is a long period comet with an orbital period of 2875 years and high orbital inclination of 70.4° to the ecliptic plane. The comet passed its perihelion on February 7.5, 2006 at a distance of 6.8 AU from the Sun. Over an extended period of time the comet had an asteroidal appearence without any sign of cometary activity. According to informal classification proposed by Jewitt (2005) it was assigned as Damocloid, an asteroidal object with orbital parameters resulting in a Tisserand's parameter $T_J \leq 2$. Later VQ94 was excluded from the list of Damocloids, which is published in JPL Small-Body Database, because cometary activity was finally detected. An effective nuclear radius measured at the asteroidal stage was found to be equal to 40.7 km assuming a red geometric albedo $p_R$=0.04 (Jewitt, 2005). The large size of the nucleus was confirmed by Weiler et al. (2011). Using the observations of the comet published by the Minor Planet Center Extended Computer Service, the authors found that the nuclear radius was equal to 52.8 km with a 27% uncertainty. Tegler et al. (2003) estimated color indices, B-V=0.92±0.04 and V-R=0.47±0.02, from the photometric observations made on December 31, 2002, when the comet was inactive.

Cometary activity of the object was first detected at the end of August 2003, when a distance from the Sun was 8.9 au. A prominent 10″ coma with a fanlike morphology spanning position angles between 180° and 300° was found on images taken by Jewitt on August 28.5 UT with the University of Hawaii 2.2-m telescope (Green, 2003; Jewitt, 2005).

Spectroscopic and photometric observations of the active comet were made with the 6-m telescope SAO RAS (Special Astrophysical Observatory of Russian Academy of Sciences) on March 9.9, 2006, shortly after perihelion passage (Korsun et al., 2006). An unexpected result was found: in addition to the continuum a number of strong molecular emissions, CN, $C_3$, $CO^+$, and $N_2^+$, were detected in the spectrum of the comet. Detections of the molecular emissions at large heliocentric distances are very rare events in the cometary history. At heliocentric distances larger than 5 au emissions were detected so far in a few comets: C/1961 R1 (Humason) ($CO^+$) (Dossin, 1966), 29P/Schwassmann–Wachmann 1 ($CO^+$, CN) (Cochran et al., 1980; Cochran et al., 1982; Cochran and Cochran, 1991; Cochran et al., 1991), C/1995 O1 (Hale-Bopp) (CN, $C_3$) (Rauer et al., 2003), centaur Chiron (CN) (Bus et al., 1991), and at last our data on C/2002 VQ94 (LINEAR) (CN, $C_3$, $CO^+$, $N_2^+$) (Korsun et al., 2006). One year later, on April 10.0, 2007, when the comet was at a heliocentric distance of 7.33 au, we again obtained the spectroscopic and photometric data of the comet. The neutral species were not detected in the spectra, while emissions of $CO^+$ and $N_2^+$ were confidently measured (Korsun et al., 2008). A map of column densities of $CO^+$ in the coma of the comet was built from the images obtained through the $CO^+$+continuum SED415 filter and the continuum SED537 filter (Ivanova et al., 2009).

In this paper we present the results obtained from our new observations, which cover a large range of heliocentric distances, up to 16.84 au. The paper is organized as follows: observation technique and overview of the data are presented in Section 2. Section 3 describes reduction procedures applied on the spectroscopic and photometric data. Cometary spectra are presented in Section 4. Section 5 gives the details of image processing and results derived from the photometric analysis. The main findings are presented in Section 6. The section also gives a significant discussion of the reality of the $N_2^+$ detection in the cometary spectra. Finally, brief conclusions complete the paper.

2. **Observations**

Two new observing runs were carried out with the 6-m Big Telescope Alt-azimuth (BTA) operated by SAO RAS in March 2008 and March 2009. Images of VQ94 with broad band filters were obtained using the 2.5-m Nordic Optical Telescope (NOT) (La Palma, Spain) in June 2011 and July 2013.

A focal reducer SCORPIO attached to the prime focus of the BTA telescope (Afanasiev and Moiseev, 2005) was used to obtain both spectra and images of the comet. An EEV-42-40 CCD chip of 2048×2048 pixels has a full field of view of 6.1′×6.1′ with an image scale of 0.18″/pix. The transparent grisms VPHG1200B (March 13/14, 2008) and VPHG550G (March 30/31, 2009) were used as dispersers in the spectroscopic mode and a long-slit mask with 6.1′×1.0″ dimensions was projected on the cometary coma. The spectral resolution of the spectra was defined by the width of the slit and was about 5 Å and 10 Å in the 2008 and 2009 observing runs respectively. On March 13/14, 2008 the images of the comet were obtained through the narrow-band SED537 filter ($\lambda c$=530.9nm, FWHM=16.9nm), which was used as a continuum band. Rc filter with an effective wavelength of 650 nm was used on March 30/31, 2009. In order to improve the signal/noise ratio of the measured signal, on-chip binning of 2×2 and 2×1 (spatial direction) was applied to the photometric and spectroscopic frames respectively. The telescope was tracked on the comet to compensate its apparent movement during the exposures.

The spectra and images of the spectrophotometric standard stars HZ44 (March 13/14, 2008) and BD+28d4211 (March 30/31, 2009) were obtained to fulfill absolute calibration (Oke, 1990). Observations of the morning sky through the used filters were also made to provide flat-field corrections.

We continued monitoring of VQ94 with the 2.5-m Nordic Optical telescope in June 2011 and July 2013. Images were obtained with the Andalucia Faint Object Spectrograph and Camera (ALFOSC) used in imaging mode. In this mode the field of view is 6.4′x6.4′ with a 2k×2k CCD detector having 13.5 μm pixel size, corresponding to 0.19″ on the sky. The observations were conducted with R, V and B broadband filters and a field with standard stars (SA110) was also observed at different airmasses to provide information for the photometric reduction (Landoldt, 1992). Three nights could be considered as photometric with a seeing value being about 1″. Table 1 and Table 2 contain the journal of the observations performed with the 6-m BTA and NOT respectively. Geometric circumstances of the comet at the moment of the observations are also included in the tables. Variation of the appearance of the comet during the monitoring is displayed in Fig. 1.

Table 1. Log of the observations of VQ94 performed with the BTA telescope. Δ: geocentric distance, r: heliocentric distance.

| Start of exposure (UT date) | Exp. time (s) | Δ (au) | r (au) | phase angle (°) | airmass | data type | filter / spectral range (Å) |
|---|---|---|---|---|---|---|---|
| March 13.9364, 2008 | 1200 | 7.62 | 8.36 | 4.81 | 1.23 | spectrum | 3600-5400 |
| March 13.9510, 2008 | 1200 | 7.62 | 8.36 | 4.81 | 1.20 | spectrum | 3600-5400 |
| March 13.9654, 2008 | 1200 | 7.62 | 8.36 | 4.81 | 1.17 | spectrum | 3600-5400 |
| March 13.9899, 2008 | 300 | 7.62 | 8.36 | 4.81 | 1.14 | image | SED537 |
| March 30.9699, 2009 | 10 | 8.96 | 9.86 | 2.54 | 1.42 | image | R |
| March 30.9735, 2009 | 10 | 8.96 | 9.86 | 2.54 | 1.42 | image | R |
| March 30.9752, 2009 | 10 | 8.96 | 9.86 | 2.54 | 1.42 | image | R |
| March 30.9767, 2009 | 10 | 8.96 | 9.86 | 2.54 | 1.43 | image | R |
| March 30.9896, 2009 | 900 | 8.96 | 9.86 | 2.54 | 1.44 | spectrum | 3100-7300 |
| March 31.0007, 2009 | 900 | 8.96 | 9.86 | 2.54 | 1.47 | spectrum | 3100-7300 |
| March 31.0117, 2009 | 900 | 8.96 | 9.86 | 2.54 | 1.54 | spectrum | 3100-7300 |
| March 31.0229, 2009 | 900 | 8.96 | 9.86 | 2.54 | 1.59 | spectrum | 3100-7300 |

Table 2. Log of the observations of VQ94 performed with the NOT. Δ: geocentric distance, r: heliocentric distance.

| Middle of exposure | number of exposures | total exposure time (s) | Δ (au) | r (au) | phase angle (°) | airmass | filter |
|---|---|---|---|---|---|---|---|
| June 03.9670, 2011 | 2 | 240 | 12.59 | 13.40 | 2.68 | 1.69 | V |
| June 03.9675, 2011 | 5 | 600 | 12.59 | 13.40 | 2.68 | 1.67 | R |
| June 03.9810, 2011 | 2 | 360 | 12.59 | 13.40 | 2.68 | 1.71 | B |
| June 04.9655, 2011 | 10 | 1800 | 12.60 | 13.40 | 2.74 | 1.70 | R |
| June 04.9656, 2011 | 2 | 360 | 12.60 | 13.40 | 2.74 | 1.66 | V |
| June 04.9882, 2011 | 2 | 360 | 12.60 | 13.40 | 2.74 | 1.75 | B |
| July 06.8979, 2013 | 1 | 300 | 16.42 | 16.84 | 3.18 | 2.54 | B |
| July 06.9018, 2013 | 3 | 900 | 16.42 | 16.84 | 3.18 | 2.56 | R |
| July 06.9058, 2013 | 1 | 300 | 16.42 | 16.84 | 3.18 | 2.63 | V |

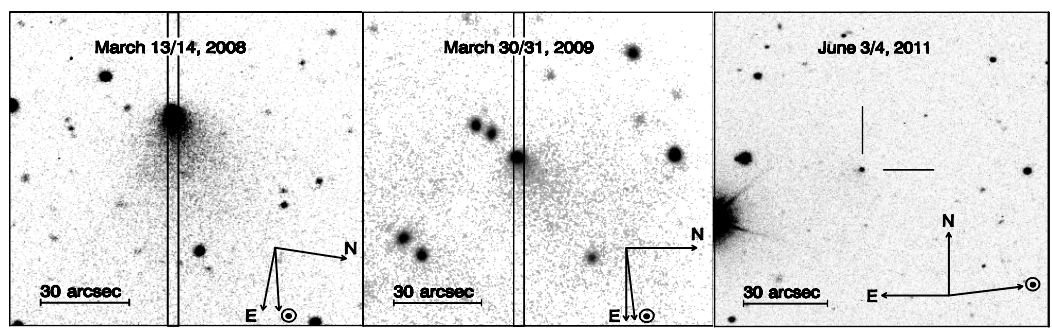

Figure 1. 120″×120″ extractions from the summarized images of VQ94. North, East, sunward direction, and scale bar are indicated. The vertical narrow box across the comet marks the position of the slit in the spectroscopic mode.

## 3. Data reductions

In order to reduce long-slit spectra the raw spectral data obtained with the BTA were processed using program codes developed in SAO RAN. Each recorded spectrogram was processed separately. An averaged bias frame was removed from the data. An exposure of the lamp with smoothly varying energy distribution was used to compensate a nonuniform sensitivity of the chip's pixels (flat-field). To do wavelength calibrations we exposed a He-Ne-Ar-filled lamp. The wavelength scale was linearized in order to facilitate data interpretation. Night sky spectrum was removed using its level measured in each column over the zones free of the cometary coma. Photometric calibration of the observed spectra was provided using the observed transformation of the standard star spectrum. Spectral behavior of atmospheric extinction was taken from Kartashova and Chunakova (1978). To increase the signal/noise ratio, we combined individual spectra by median filtering between the frames. This also removed the cosmic ray events. The resulting spectra were collapsed in the spatial direction when a coma was detected.

The images of the comet and standard stars obtained through the SED537 and Rc filters were bias subtracted and subsequently flat-fielded using twilight sky images. The background signal was removed from the images, and the technique of aperture photometry was applied to integrate the fluxes from the comet and standard stars. A number of apertures with radii from 4 to 40 pixels were used consequently to build surface brightness profiles of the comet. The smallest aperture size was chosen to be equal approximately to one FWHM of a star photometric profile; the bigger aperture size had to be big enough to include the signal from the whole coma. Convolution of the filter transmission curves with the spectra of the standard stars and the Sun gave their effective magnitudes for the filter bands (Arvesen et al., 1969; Oke, 1990). Since the standard stars were not observed over the wide interval of airmasses, we evaluated the steadiness of the sky analyzing 'seeing' with the stars' photometric profiles. The field stars were chosen in the individual frames, which were obtained in the course of the observation of the comet, to extract the FWHM and analyze its possible instability. The comet observations lasted approximately for 3 hours in both 2008 and 2009. In 2008 the stars' profiles started to change gradually after 2 hour period of the observation but the standard star images had been already recorded. It was estimated that in 2008 the 'seeing' slightly varied but was still better than 1.7″. In 2009 the atmospheric conditions, which provided 'seeing' around 1.5″, was relatively steady. The differences between the air masses of the comet and standard stars were about 0.1 in 2008 and about 0.3 in 2009. In order to compensate the difference in the extinction, the monochromatic extinction coefficients measured for Pik Terskol Observatory, which is located at the same mountain area at the altitude of 3100m, were used (Kulyk et al., 2004).

For the NOT observational data we first pre-processed the data with an averaged bias image and with averaged twilight flat fields, which were obtained for each filter and for the two nights of June 2011 and the one of July 2013. The different standard star images were pre-processed in a similar way and complete photometric reduction (i.e., the computation of the different photometric coefficients: extinction coefficient, color term and system zero point for each filter) was fulfilled using these images for each night with the NOAO IRAF reduction package.

## 4. Spectroscopy

Qualitative inspection of the spectrum obtained in March 2008 shows that the emissions observed during our previous observing period are still present. To separate the emission and continuum components we fitted the continuum with a solar spectrum modified along dispersion due to the reddening effect. The solar spectrum was taken from Neckel and Labs (1984). It was degraded to the resolution of our observations by convolving the solar spectrum with the appropriate instrumental profile and normalized to the flux from the comet around 5000 Å. The resulting solar spectrum was multiplied by a polynomial depicted in Fig. 2B. Fig. 2A shows a derived continuum, which is superimposed on the observed spectrum.

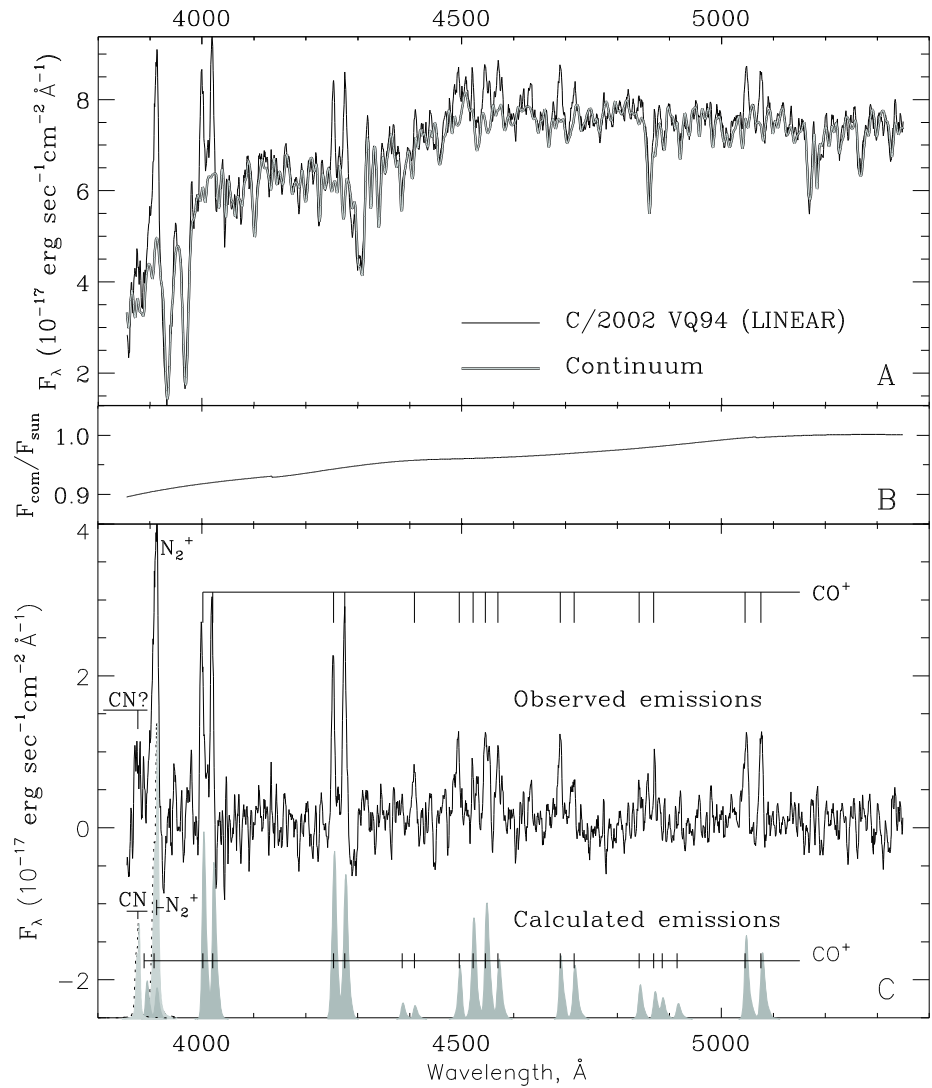

Figure 2. Spectrum of VQ94 observed at a heliocentric distance of 8.36 au: A. – the observed spectrum with the superimposed continuum; B – the polynomial correction of the solar spectrum; C – the emission component.

The emission component was obtained by subtracting the calculated continuum from the observed spectrum. It is depicted in Fig. 2C. The emissions of $CO^+$ dominate in the spectrum. Their assignments were made using the Ultraviolet and Visible Spectroscopic Database for Atoms and Molecules in Celestial Objects compiled by Kim (1994). The calculations were made assuming the Boltzmann distribution of the level populations and the result is displayed at the bottom of Fig. 2C. We assigned the (4, 0), (3, 0), (2, 0), (1, 0), (3, 1), (2, 1), (4, 2), (0, 0), and (1, 1) vibrational transitions of the $A^2\Pi$–$X^2\Sigma$ band system (the Comet-tail bands) of $CO^+$. Using the LIFBASE[1] software (Luque and Crosley, 1999) we calculated the spectra of CN and $N_2^+$, which we considered as potential targets (Fig. 2C, bottom). While assignment of CN is questionable, the $N_2^+$ emission is undoubtedly present in the cometary spectrum. More details of our assignments are in Table 3, where we listed identified molecular emissions and their peak intensities.

---
[1] http://www.sri.com/cem/lifbase

Table 3 Observed emissions

| Observations | | Assignments | | | |
|---|---|---|---|---|---|
| Wavelength, Å | Peak intensity[a] | Molecule | System | Band | Wavelength, Å |
| 3913.5 | 4.13 | $N_2^+$ | $B^2\Pi–X^1\Sigma$ | (0,0) | 3914.7 |
| 3999.1 | 2.71 | $CO^+$ | $A^2\Pi–X^2\Sigma$ | (3,0) | 4001.8 |
| 4019.2 | 3.04 | $CO^+$ | $A^2\Pi–X^2\Sigma$ | (3,0) | 4121.2 |
| 4252.7 | 2.27 | $CO^+$ | $A^2\Pi–X^2\Sigma$ | (2,0) | 4253.3 |
| 4274.6 | 2.91 | $CO^+$ | $A^2\Pi–X^2\Sigma$ | (2,0) | 4275.1 |
| 4408.9 | 0.84 | $CO^+$ | $A^2\Pi–X^2\Sigma$ | (3,1) | 4408.3 |
| 4493.0 | 1.27 | $CO^+$ | $A^2\Pi–X^2\Sigma$ | (4,2) | 4494.9 |
| 4520.4 | 0.92 | $CO^+$ | $A^2\Pi–X^2\Sigma$ | (4,2) | 4521.8 |
| 4546.0 | 1.26 | $CO^+$ | $A^2\Pi–X^2\Sigma$ | (1,0) | 4546.5 |
| 4570.1 | 1.09 | $CO^+$ | $A^2\Pi–X^2\Sigma$ | (1,0) | 4570.3 |
| 4690.0 | 1.24 | $CO^+$ | $A^2\Pi–X^2\Sigma$ | (2,1) | 4689.0 |
| 4717.0 | 0.63 | $CO^+$ | $A^2\Pi–X^2\Sigma$ | (2,1) | 4716.2 |
| 4841.2 | 0.63 | $CO^+$ | $A^2\Pi–X^2\Sigma$ | (0,0) | 4841.9 |
| 4870.8 | 1.04 | $CO^+$ | $A^2\Pi–X^2\Sigma$ | (0,0) | 4870.7 |
| 5048.3 | 1.26 | $CO^+$ | $A^2\Pi–X^2\Sigma$ | (1,1) | 5045.9 |
| 5077.1 | 1.27 | $CO^+$ | $A^2\Pi–X^2\Sigma$ | (1,1) | 5077.1 |

[a] $\times 10^{-17}$ ergs s$^{-1}$ cm$^{-2}$ Å$^{-1}$

Having the flux integrated within the observed ionic emission features, we can derive the average column densities of the molecular emissions $N$ using the expression:

$$N = \frac{4\pi F}{\Omega g}, \qquad (1)$$

where $F$ is the integrated molecular band flux, $g$ is the fluorescence efficiency, or $g$-factor, of the molecule, which was scaled to the appropriate heliocentric distance according to $g(r) = g(1\ au) * r^{-2}$, and $\Omega$ is the solid angle subtended by the spectrograph slit at the cometary coma. Using fluorescence efficiencies at 1 au of $7.0 \times 10^{-2}$ photons s$^{-1}$ mol$^{-1}$ for the $N_2^+$ (0,0) band (Lutz et al., 1993) and $3.85 \times 10^{-3}$ photons s$^{-1}$ mol$^{-1}$ for the CO+ (2,0) band (Magnani and A'Hearn, 1986) and integrating fluxes within the above mentioned bands we calculated the average column densities as $2.04 \times 10^9$ mol cm$^{-2}$ for $N_2^+$ and $3.26 \times 10^{10}$ mol cm$^{-2}$ for $CO^+$. The possible contamination of the $N_2^+$ emission by the $CO^+$ (5,1) emission was taken into account. We estimated that its contribution to the total signal does not exceed 7%. The derived values of the column density yield $N_2^+/CO^+=0.06$ ratio measured within the projected slit.

The same data processing was used to fit the continuum observed in March 2009. The superimposed calculated and observed continua are shown in Fig. 3A. The corresponding polynomial correction is depicted in Fig. 3B. Any traces of the emissions, which were clearly identified in the spectrum obtained one year before (see Fig. 2C) are absent.

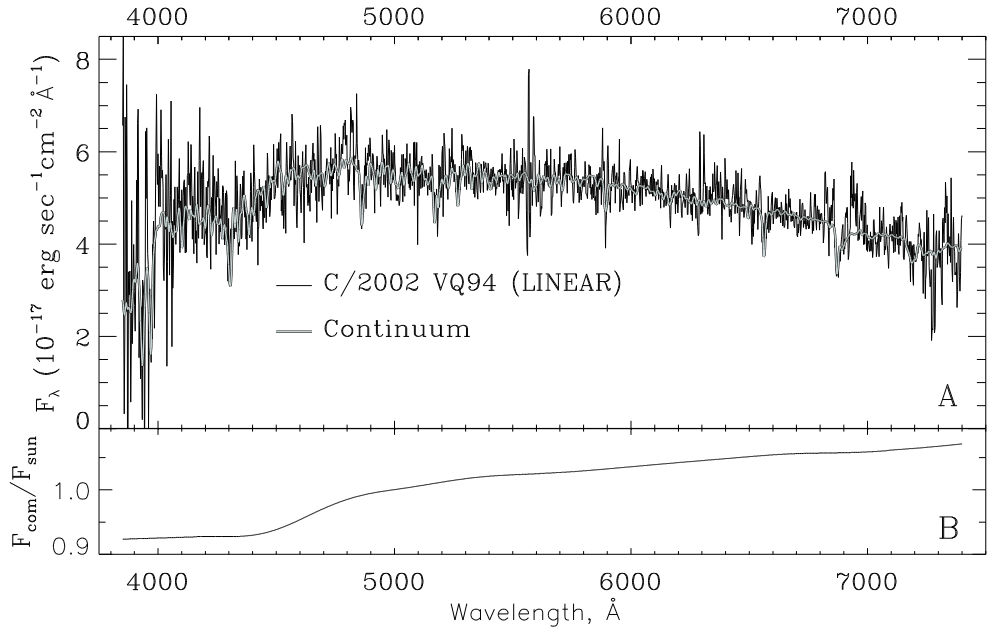

Figure 3. Spectrum of VQ94 observed at a heliocentric distance of 9.86 au: A – the observed spectrum with the superimposed continuum; B – the polynomial correction to the solar spectrum.

## 5. Image analysis

At the beginning all the images were analyzed to inspect the image morphology. The images obtained in 2008 and 2009 revealed clearly the presence of the coma. The images of the comet obtained with the NOT in 2011 showed a very faint coma and there was no sign of the cometary activity on the images from 2013. Therefore, for the data obtained with the NOT we first extracted subimages centered on the target and a field star and coadded them separately for each filter and each night (B, V and R). From these images using the photometric coefficients computed with the standard star observations (see above) we derived radial profiles for both the target and the reference star (the background signal was previously removed). The radial profiles obtained for the reference star were then shifted in magnitude to adjust their maximum to the corresponding target profiles (same night and same filter). It was then possible to see clearly the presence of the coma for the data obtained in June 2011. Fig. 4 presents the results obtained with the R filter for the night 4 June 2011.

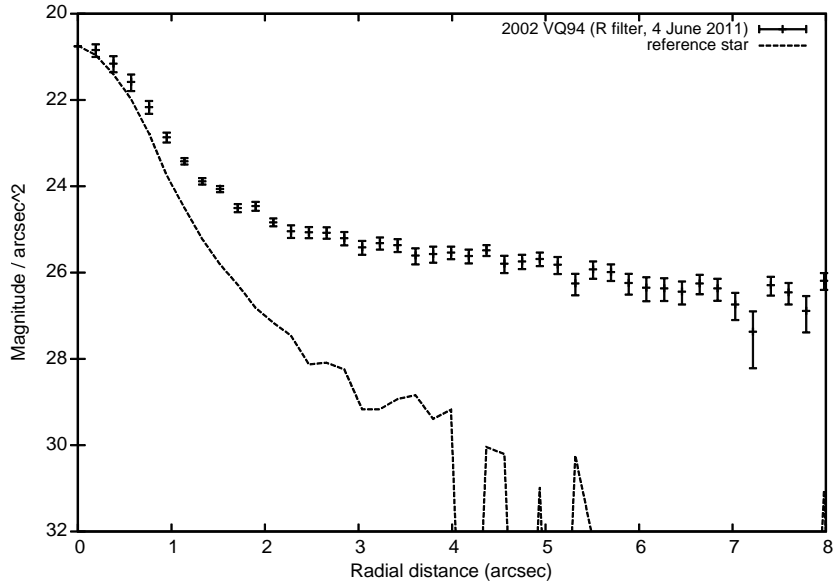

Figure 4. Radial profile computed for VQ94 in the R band (4 June 2011). It is represented with a stellar radial profile (solid line) adjusted in intensity for comparison with the PSF. The excess of intensity of VQ94 corresponds to the coma.

The images obtained on 3 June 2011 were finally not taken into account because of the presence of a small artifact in the images, preventing the computation of a good radial profile of the target. Comparison of the radial profiles of the comet image obtained in 2013 with the stellar profile showed no difference.

We used the radial profiles to derive $Af\rho$ values. This parameter was introduced by A'Hearn et al. (1984) to measure cometary activity in such a way that it does not differ too much from one observer to another and to be independent of unknown parameters, such as grain albedo or grain size. It is a product of the albedo ($A$), the filling factor ($f$), i.e., the ratio of the total cross section of grains within the field of view by the area of the field of view and $\rho$, the linear radius of the field of view at the comet distance.

For a given band it is possible to compute it from the magnitude of the Sun in the same band. We can write:

$$Af\rho = [2.4686 \times 10^{19} r^2 \Delta 10^{0.4(M_{Sun} - M)}]/\alpha \qquad (2)$$

where $r$ is the heliocentric distance (au), $\Delta$ – the geocentric distance (au), $M_{Sun}$ – the magnitude of the Sun (−26.09, −26.74, and −27.26 for the B, V, R – bands respectively, −26.98 for the Rc filter, and −26.66 for the SED537 filter), $M$ – the overall magnitude of the coma and $\alpha$ – the apparent diameter of the field of view expressed in arcsec. The resulting $Af\rho$ value is expressed in cm and it is supposed to be, more or less, independent of the aperture size used to measure it (if the radial profile follows a $\rho^{-1}$ dependence).

The difficulty in this formula is to compute the overall coma magnitude, $M$. The fastest way to estimate it is to do aperture photometry on the target with a relatively large diameter (a few arcsec). For very active comets it can provide valuable results because the brightness of the nucleus, compared to the coma, is small. This is the case for the VQ94 images obtained in March 2008. Previously estimated effective radius of the cometary nucleus of about 50 km along with the adopted albedo value of 0.04 allows us to estimate the contribution of flux from the bear nucleus. The measured magnitude of the comet within the circular aperture of about 6″ (approximately 3×FWHM of a radial profile of a field star) points out that the contribution of the flux from the cometary coma is probably comparable to that of the bare nucleus. For the images obtained in 2009 and especially in 2011, because of the faintness of the coma, it was necessary to take care of the light that comes from the nucleus. From the radial profile of the comet we subtracted the flux due to the nucleus, which was supposed to be close to the stellar profile adjusted in intensity to the maximum of brightness. Fig. 5 presents the results for the data obtained on 4 June 2011 for the three different filters.

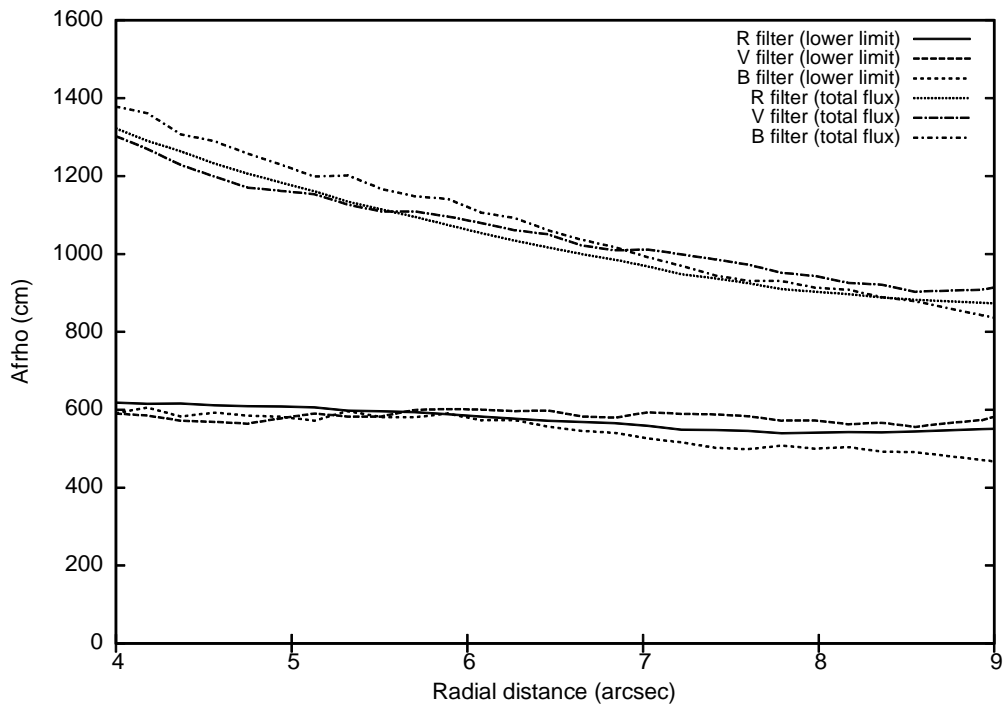

Figure 5. $Af\rho$ values computed from the NOT coadded images obtained on 4 June 2011. The upper lines represent $Af\rho$ computed using all the flux present in a disk with a radius corresponding to the radial distance of the x-axis. The lower lines indicate $Af\rho$ obtained after subtraction of the stellar profile adjusted in intensity with the maximum of brightness.

It can be seen in this figure that the $Af\rho$ values are, more or less, constant for different radial distances when the nucleus contribution has been subtracted, which is expected for a coma having a $\rho^{-1}$ dependence. The radial distance range represented in the figure corresponds to distances, where the subtracting of the nucleus contribution is not too critical – above about 4″ – and where no stars disturb the results – above about 9″. The $Af\rho$ values obtained, when taking into account the whole flux, are also represented, and, as it can be seen, there is a strong dependence on the radial distance. The real $Af\rho$ values are probably more important than the ones computed after subtraction of the nucleus contribution, because even at the maximum of brightness there is a contribution from the coma. Therefore the real $Af\rho$ values can be estimated by averaging the two extreme limits. From Fig. 5 we get $Af\rho$=800 cm for the B, V, and R band for 4 June 2011 with the lower and upper limits on $Af\rho$ being at 600 cm and 1000cm.

A similar data processing was done for the BTA images obtained on 30 March 2009. Fig. 6 and Fig. 7 present the results obtained from these data after coaddition of 4 different images obtained on 30 March 2009 for the radial profile and the derived $Af\rho$ values respectively.

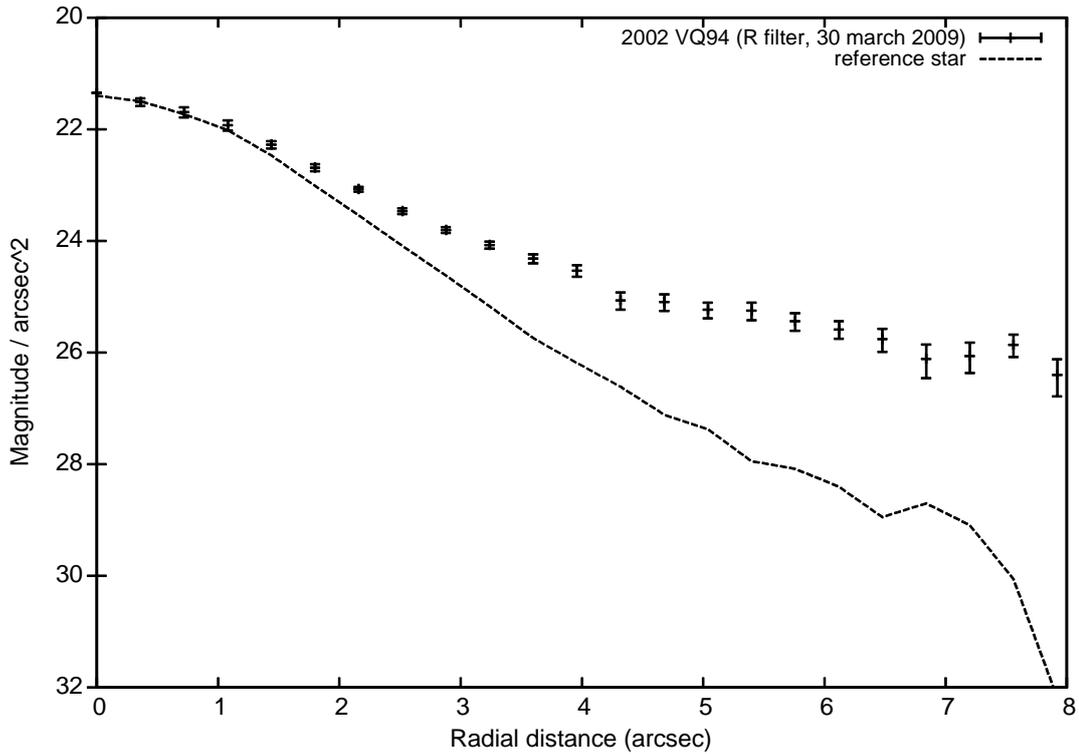

Figure 6. Radial profile of VQ94. The coadded images were obtained with the BTA on March 30, 2009 (Rc filter). It is represented with a stellar radial profile (solid line) adjusted in intensity for comparison with the PSF. The excess of intensity of VQ94 corresponds to the coma.

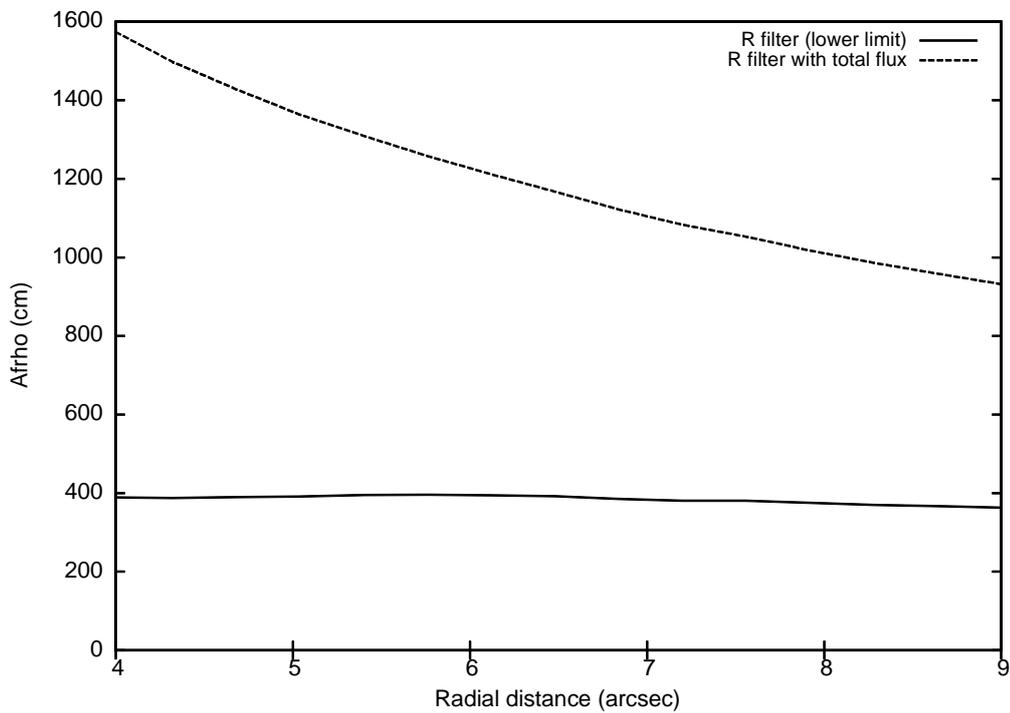

Figure 7. Dependence of $Af\rho$ values on radial distance derived from coadded images obtained on March 30, 2009 with the BTA. The upper line indicates $Af\rho$ dependence computed using all the flux present in a disk with a radius corresponding to the radial distance of the x-axis. The lower line indicates $Af\rho$ obtained after subtraction of the stellar profile adjusted in intensity with the maximum of brightness.

Finally it can be seen that a lower limit for $Af\rho$ is about 400 cm for the BTA data (for Rc filter) obtained in 2009 and the real values must be equal to 800 cm, with the lower and upper limits of 400 and 1200 cm, calculated after the subtracting of the contribution of the nuclear flux and with all flux respectively. For the data obtained in 2008, the value of $Af\rho$ amounts to 2000 cm thus indicating a considerable level of the physical activity of VQ94 at a heliocentric distance of 8.36 au. To calculate $Af\rho$ the aperture radius of 15″ was used to cover the outermost region of the cometary coma, where the signal exceeded the $3\sigma$ level of the background noise.

From the $Af\rho$ values we estimated dust production rate of the comet nucleus. For this purpose we used an approach described by Weiler et al. (2003), taking into account the discussion given by Fink and Rubin (2012). To calculate dust mass production rate we used the following equation:

$$Q_M = Q_N \left(\frac{4\pi}{3}\right) \left[\int \rho_d(a) a^3 f(a) da\right] \qquad (3)$$

Here, $Q_N$ is the dust number rate; $\rho_d(a)$ is the density of a grain, which depends on the grain radius, $a$; $f(a)$ describes a differential particle size distribution. According to Fink and Rubin (2012) the particle size distribution $f(a)$ is the critical parameter that influences the level of dust production rate. Given that we used two different functions to describe the differential particle size distribution: in the simple form of $f(a) \sim a^{-n}$ and somewhat more complex expression $f(a) \sim (1 - \frac{a_0}{a})^M \times (\frac{a_0}{a})^N$, used by Hanner (1983). In the former expression we fixed the power index $n$ at 4, following the results of the model simulation of the dust environment of some active distant comets (Korsun, 2005; Korsun et al., 2010). In the latter expression $a_o$ denotes a minimum grain radius, $M$ and $N$ are parameters, which provide a peak of the function for a particle size of $a_m = a_o \times [\frac{M+N}{N}]$. We used $M=1.8$ and $N=3.6$, which correspond to the maximum of the function for grains of about 15μm. Fixing the minimum of a grain radius at 5μm, we took into account the previous findings leading to the conclusion that comae of distant comets are probably populated by large particles, which have low outflow velocities (Korsun et al., 2010; Meech et al., 2009). $\rho_d$, the size dependent density of dust particles, was chosen in accordance with Newburn and Spinrad (1985), as $\rho_d(a) = \rho_0 - \rho_l \times [\frac{a}{a+\tilde{a}}]$, where $\rho_0$ is 3000 kg m$^{-3}$, $\rho_l$ is 2200 kg m$^{-3}$, and $\tilde{a}=2$μm.

$Q_N$ was calculated with the following expression (Jorda, 1995):

$$Q_N = Af\rho[(2\pi^2 A_B \Phi(\alpha))]^{-1} \left[\int (f(a) a^2)/v(a) da\right]^{-1} \qquad (4)$$

In this expression $A$ is a Bond-albedo of a particle, the value of 0.1 was assumed, $\Phi(\alpha)$ is a phase function that was adopted in the form of $10^{-0.4 \times 0.04 \times \alpha}$ because the observations were made under small phase angles between 4.80 and 2.54°. For the calculation of the dust particles velocities the expression $v(a) \sim r^{-0.5} \times a^{-0.5}$ was adopted, where $a$ is expressed in μm and velocities in m s$^{-1}$, $r$ is the heliocentric distance in au (Korsun et al., 2010). We took minimum and maximum of the particle radii to be equal to 5 μm and 1000 μm as limits for the integration procedure.

The dust mass production rate and dust number rate for the two different particle size distributions are presented in Table 4.

Table 4. Dust mass production rate for comet C/2002 VQ94

| Differential particle size distribution | Dust mass rate/ Dust number rate | 2008 | 2009 | 2011 |
|---|---|---|---|---|
| $f(a) \sim a^{-4.0}$ | $Q_M$ (kg s$^{-1}$) | ~11 | ~4 | ~3 |
|  | $Q_N$ (s$^{-1}$) | ~1×10$^{12}$ | ~5×10$^{11}$ | ~4×10$^{11}$ |
| $f(a) \sim (1 - \frac{a_0}{a})^M \times (\frac{a_0}{a})^N$  M=1.8, N=3.6, $a_o$=5μm | $Q_M$ (kg s$^{-1}$) | ~18 | ~6 | ~5 |
|  | $Q_N$ (s$^{-1}$) | ~1×10$^{11}$ | ~5×10$^{10}$ | ~4×10$^{10}$ |

The *Afρ*, calculated from the images of the comet at a distance of 8.36 au, indicated a high level of the physical activity of the cometary nucleus as compared to some other comets at approximately the same distances from the Sun: the *Afρ* of about 1600 cm and 600 cm were estimated for comets C/2001 G1 (LONEOS) and C/2001 T4 (NEAT) at distances of 8.3 and 8.6 au respectively (Hicks et al., 2007). Above a heliocentric distance of 9 au VQ94 probably lowered the activity significantly. Although it was still high (*Afρ* ≥ 400 cm), but was considerably lower than that measured for comet C/1995 O1 (Hale-Bopp) or C/2007 D1 (LINEAR) at similar heliocentric distances (Weiler et al., 2003; Epifani et al., 2006). However, VQ94 remained still active even at such large distance as 13.40 au from the Sun. However, as our observations revealed, the comet did not show any sign of the activity at a distance of 16.84 au.

Finally, the images obtained from this last observing run were used to estimate the color of the inactive nucleus and its radius. The B-V and V-R colors were found to be equal to 1.07±0.05 and 0.54±0.03. V-R value confirms the earlier measurements, when the comet was inactive yet on the inbound part of its orbit (Jewitt, 2005; Tegler, 2003;) But B-V exceeds slightly the previously reported results. This discrepancy is possibly caused by underestimating of the extinction in the B band, since the observations were made at the airmasses greater than 2.5.

The radius of VQ94 was estimated with an equation taken from Russel (1916) assuming the albedo value of 0.04 and phase coefficient of 0.032. The V magnitude of 21.50±0.04 corresponds to the radius of 48±2 km that is close to the value of 52.8 km reported by Weiler et al. (2011).

## 6. Discussion

The important result of the spectroscopic observations of comet VQ94 presented in this paper is the detection of well-developed ionic emissions of CO$^+$ and N$_2^+$ at a record heliocentric distance, 8.36 au. At such large heliocentric distances, these ionic emissions have not been observed so far in any Solar system bodies at optical wavelengths. Assignment of CO$^+$ is undoubted because we detected a number of the molecular bands with expected profiles and ratios in intensities. As for the assignment of $N_2^+$, we consider it is necessary to give additional arguments in support of its identification. The reason is that there are restrictions on detection of $N_2^+$ in cometary comae. While many observers reported the detections of the $N_2^+$ emissions from the investigation of low resolution spectroscopic data of comets in the past (see, for example, Greenstein and Arpigny (1962), Lutz et al. (1993), Wyckoff and Theobald (1989)), more recent high resolution spectroscopic observations of a number of comets have not confirmed the presence any vestiges of the $N_2^+$ rotational lines (Cochran et al., 2000; 2002). Moreover, recently Mousis et al. (2012) propose a scenario that could explain an apparent $N_2$ deficiency in comets. They argue that the $N_2$ deficiency might be a consequence of poor trapping efficiency of $N_2$ and postaccretion internal heating engendered by the decay of radiogenic nuclides. We consider that the scenario can be valid for comets where detection of $N_2^+$ was really failed, but it is bounded by a number of specific restrictions and does not cover all possible temperatures and chemical singularities of the earlier Solar system.

We made the spectroscopic observations of the comet during the astronomical night and at low airmasses. This significantly reduces the risk of detection of telluric emissions of $N_2^+$. Our spectra were recorded in long-slit mode. The length of the slit was 6.1′ and projection of the cometary coma on the slit was about 70″. Thus, we can accurately subtract spectrum of the night sky, and, at the same time, check the presence of possible telluric and/or auroral emissions of $N_2^+$. We did not register any emission features in this spectral region in the spectrum of the night sky, which shows that the discussed emission feature originated from the cometary coma of VQ94. We also observed spectroscopically the distant comet C/2007 D1 (LINEAR) the same night and did not find any traces of $N_2^+$ in the spectrum.

A validation that the emission feature is just the $N_2^+$ emission was made using a calculated spectrum of $N_2^+$. In Fig. 8 we show a section of the observed spectrum with a superimposed theoretical spectrum of $N_2^+$. The later was built using the LIFBASE software (Luque and Crosley, 1999), assuming the Boltzmann distribution of the level populations at 140 K. Being reduced to the spectral resolution of the cometary spectrum it fits well the observed spectrum. The $CO^+$ emissions, which fall in the spectral region, were calculated with The Ultaraviolet and Visible Spectroscopic Database for Atoms and Molecules in Celestial Objects (Kim, 1994) and also presented in the figure. It can be seen that the $CO^+$ emissions cannot fit the observations because they are two-peaked and distinctly weaker.

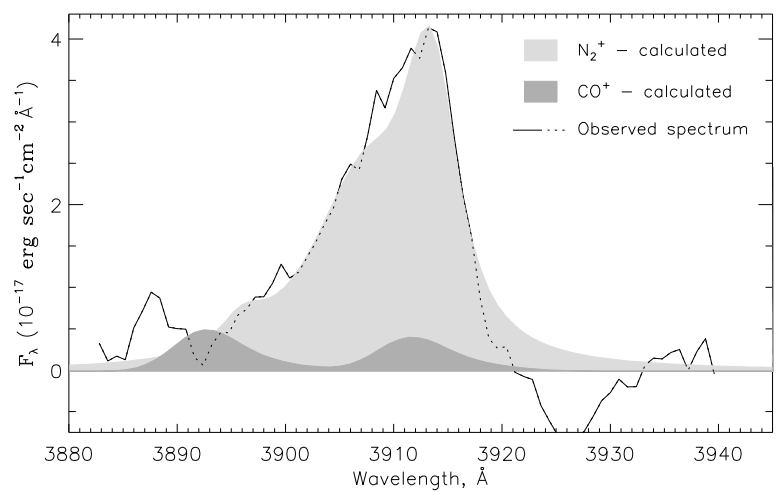

Figure 8. Section of the observed spectrum fitted by the $N_2^+$ theoretical spectrum and the $CO^+$ theoretical spectrum.

And finally, as we observed the long slit spectra, we computed the spatial profiles of the identified emissions. To derive the distribution of the flux across the slit we performed integration of the two dimensional spectrum along dispersion within the appropriate spectral windows: 3895 – 3919 Å for the $N_2^+$+continuum spatial profile and 3947 – 3961 Å for the $N_2^+$ underlying continuum; 3992 – 4024 Å for the $CO^+$(3-0)+continuum spatial profile and 4035 – 4074 Å for the $CO^+$ underlying continuum. True levels of the corresponding underlying continua were calculated using the Flux(3895:3919)/Flux(3947:3961) and Flux(3992:4024)/Flux(4035:4074) ratios obtained from the fitted cometary continuum (see Fig. 2A). Fig. 9 demonstrates the distribution of the total flux (emission+continuum) along the slit, the calculated continuum and the detached $CO^+$ and $N_2^+$ spatial profiles. It is clearly seen that the spatial profiles of $CO^+$ and $N_2^+$ are similar. They are very asymmetric and isolated within the observed coma. Their maxima are shifted with respect to the continuum maximum about of 2″ in the tailward direction. Therefore, we conclude that the $N_2^+$ as well as $CO^+$ emissions features belong to the comet.

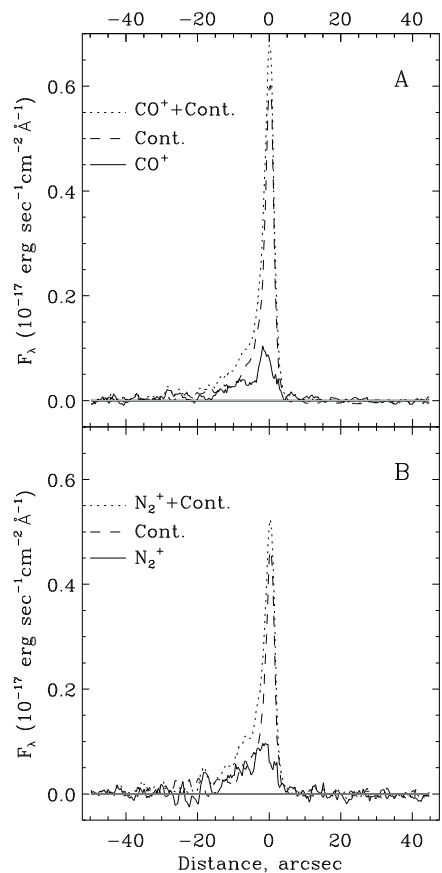

Figure 9. The spatial profiles of the identified gaseous emissions with the underlying continua, the underlying continua, and the detached cometary $CO^+$ and $N_2^+$.

The derived value of $N_2^+/CO^+$=0.06 is in agreement with the compiled data for comets, for which $N_2^+$ and $CO^+$ were measured (Cochran et al., 2000). The result somewhat differ from our previous data (Korsun et al., 2006; 2008). However, taking into account accuracy of the subtraction of the continuum, possible contaminations of the $N_2^+$ band and the fact that the ratio was measured in different regions of the coma the reality of the variation is questioned.

The spectrum of the comet is characterized by noticeable predominance of the $CO^+$ and $N_2^+$ emissions (Korsun et al., 2006; 2008; this work). It is known from the previous history that the similar spectroscopic features have been detected in the spectra of comets C/1908 R1 (Morehouse) (de La Baume Pluvinel and Baldet, 1911), C/1961 R1 (Humason) (Greenstein, 1962), and 29P/Schwassnann-Wachmann 1 (Korsun et al., 2008). Meanwhile, Cochran et al. (2000; 2002) confirm identification of $CO^+$ only in the spectra of comet 29P/Schwassnann-Wachmann 1. The above mentioned comets have quite different orbits. Comet C/1908 R1 (Morehouse) is classified as a hyperbolic comet with a perihelion at 0.95 au and an orbital inclination of 140.2°, comet C/1961 R1 (Humason) is classified as a comet not matching any defined orbit class with a perihelion at 2.13 au and an orbital inclination of 153.3°, comet 29P/Schwassnann-Wachmann 1 is classified as a Jupiter family comet with a perihelion at 5.74 au and an orbital inclination of 9.4°, comet VQ94 is classified as a comet not matching any defined orbit class with a perihelion at 6.80 au and an orbital inclination of 70.5°.

While dynamical behavior of these comets does not testify their common history, the similar spectroscopic features may indicate their origin under similar physical conditions in the Solar nebula.

Since CO and $N_2$ are most likely parent molecules of the observed ions, the physical conditions at the places of births of these comets in the early Solar system had to be favorable enough to allow formation of the CO and $N_2$ reach ices. As laboratory experiments demonstrated, the CO and $N_2$ ices, can accumulate at ~25 K (Bar-Nun et al., 2007; Notesco and Bar-Nun, 2005; Notesco et al., 2003).
.

## 7. Conclusions

We present the results of the spectral and photometric observations of comet C/2002 VQ94 (LINEAR) conducted in 2008 and 2009 with the 6-m telescope of SAO RAS (March, 2008 and March 2009) and with the 2.5-m Nordic Optical Telescope (June 2011 and July 2013). During the period covered with the observations, the comet was at heliocentric distances of 8.36, 9.86, 13.41, and 16.84 au. Both spectra and photometric images of the comet were obtained with the 6-m telescope, while the 2.5-m NOT telescope was used in the photometric mode only.

The analysis of the cometary spectra obtained in 2008 revealed the presence of the emissions of $CO^+$ and $N_2^+$, which have already been recorded in the spectra of the comet in 2006 and confirmed in 2007 (Korsun, 2006; 2008). The spectra obtained in 2009 showed no emission lines above continuum at a distance of 9.86 au from the Sun.

The analysis of the photometric images confirmed the high physical activity of the comet in 2008: the $Af\rho$ value amounted to 2000 cm. According to the photometric analysis, the activity was probably reduced after the comet passed a heliocentric distance of about 9 au. The measured $Af\rho$ values were equal to 800 cm (with the lower and upper limits of 400 cm and 1200 cm) at 9.86 au and to 800 cm (with the lower and upper limits of 600 and 1000 cm) at 13.40 au. No coma was unambiguously detected at 16.84 au. The estimated dust mass production rates from the $Af\rho$ fell in the range of 10–20 kg s$^{-1}$, 4–6 kg s$^{-1}$, and 3–5 kg s$^{-1}$ in 2008, 2009 and 2011 respectively, depending on the chosen expression for the differential particle size distribution. Although the physical activity of the cometary nucleus was probably lowered between first and second observing periods, it is worth to note that the comet still remained active even at a distance of 13.40 au.

An obvious correlation was noted between decrease of the dust production rate of the nucleus and the disappearance of the emissions in the spectrum of C/2002 VQ94 (LINEAR) at heliocentric distances greater than 9 au.

The colors of the VQ94 nucleus were estimated from the images obtained during the late stage, when the activity has probable ceased, at a distance of 16.84 au.

The V-R color index does not differ significantly from one measured at early stage of the cometary activity, when the comet entered the solar system being at a distance of 8.84 au. The B-V color index exceeds slightly the previous measured value, which is probably caused by the unfavorable geometrical condition of the observations made in 2013.

The estimated size of VQ94 nucleus was found to be 48±2 km, that is in agreement with the previous results.

**Acknowledgements**

The results are based on the observations made with 6-m telescope of SAO RAN and with the Nordic Optical Telescope, operated on the island of La Palma jointly by Denmark, Finland, Iceland, Norway, and Sweden, in the Spanish Observatorio del Roque de los Muchachos of the Instituto de Astrofisica de Canarias. The research leading to these results has thus received funding from the European Community's Seventh Framework Programme (FP7/ 2007-2013) under grant agreement No. RG226604 (OPTICON). This work was also supported by State Agency for Science, Innovation and Informatization (M/6 2013 agreement, Ukraine) and Campus France in the frame of a French–Ukrainian project "PHC Dnipro". The authors thank Anita Cochran and the anonymous reviewer for valuable comments on the manuscript.